\documentclass[12pt]{elsart}
\usepackage{amsmath,amsfonts,amssymb}
\usepackage{euscript,bbold}
\usepackage{array}
\newcommand{\sB}{\mathsf{B}}
\newcommand{\sC}{\mathsf{C}}
\newcommand{\tS}{\mathbb{S}}
\newcommand{\eB}{\EuScript{B}}
\newcommand{\eF}{\EuScript{F}}
\newcommand{\eL}{\EuScript{L}}
\newcommand{\eH}{\EuScript{H}}
\newcommand{\1}{\mathbb{1}}
\renewcommand{\d}{\partial}
\newcommand{\so}{\mathrm{so}}
\newcommand{\SO}{\mathrm{SO}}

\newcommand{\Spin}{\mathrm{Spin}}
\renewcommand{\Sp}{\mathrm{Sp}}
\newcommand{\sell}{\mathrm{s}\ell}
\newcommand{\SL}{\mathrm{SL}}
\newcommand{\CC}{\mathbb{C}}
\newcommand{\RR}{\mathbb{R}}
\newcommand{\HH}{\mathbb{H}}
\newcommand{\NPB}[3]{{\sl Nucl. Phys.} {\bf B#1} (#2) #3}
\newcommand{\CMP}[3]{{\sl Comm. Math. Phys.} {\bf #1} (#2) #3}
\newcommand{\PLB}[3]{{\sl Phys. Lett.} {\bf #1B} (#2) #3}
\newcommand{\FAA}[3]{{\sl Func. Analysis and Appl.} {\bf #1} (#2) #3}
\begin{document}
\begin{frontmatter}
\title{Supersymmetry and the cohomology of (hyper)K\"ahler manifolds}
\author[QMW,INI]{JM~Figueroa-O'Farrill \thanksref{emailjmf}
\thanksref{EPSRCj}},
\author[QMW]{C~K\"{o}hl\thanksref{emailchris}\thanksref{EPSRCc}}
and
\author[QMW]{B~Spence\thanksref{emailbill}\thanksref{EPSRCb}}
\address[QMW]{Department of Physics, Queen Mary and Westfield College,
Mile End Road, London E1 4NS, UK}
\address[INI]{Isaac Newton Institute for Mathematical Sciences, 20
Clarkson Road, Cambridge, CB3 0EH, UK}
\thanks[emailjmf]{\tt mailto:j.m.figueroa@qmw.ac.uk}
\thanks[emailchris]{\tt mailto:c.koehl@qmw.ac.uk}
\thanks[emailbill]{\tt mailto:b.spence@qmw.ac.uk}
\thanks[EPSRCj]{Supported by the EPSRC under contract GR/K57824.}
\thanks[EPSRCc]{Supported by an EPSRC postgraduate studentship.}
\thanks[EPSRCb]{EPSRC Advanced Fellow.}
\begin{abstract}
The cohomology of a compact K\"ahler (resp. hyperK\"ahler) manifold
admits the action of the Lie algebra $\so(2,1)$ (resp. $\so(4,1)$).
In this paper we show, following an idea of Witten, how this action
follows from supersymmetry, in particular from the symmetries of certain
supersymmetric sigma models.  In addition, many of the fundamental
identities in Hodge--Lefschetz theory are also naturally derived from
supersymmetry.
\end{abstract}
\end{frontmatter}

\section{Introduction}

It is a classical result going back to Hodge and Lefschetz
\cite{GriffithsHarris} that the de~Rham cohomology of a compact
K\"ahler manifold admits an action of the Lie algebra $\sell(2) \cong
\so(2,1)$.  This action is generated by the operations of exterior
product with the K\"ahler form and its adjoint operation with respect
to the Hodge inner product.  A more recent result of Verbitsky
\cite{Verbitsky-so(5),Verbitsky} states that if the manifold is
hyperK\"ahler, then the $\so(2,1)$ action is part of a larger $\so(4,1)$
action, which is now generated by exterior products with each of the
three K\"ahler forms and their adjoints.  Recently Witten
\cite{WittenTalk} has suggested that this can be understood from
supersymmetry.

It is well-known that requiring supersymmetry in a sigma model imposes
strong restrictions on the geometry of the target manifold $X$.  The
earliest result in this direction was the observation \cite{Zumino}
that the four-dimensional sigma model can be made supersymmetric if
and only if $X$ admits a K\"ahler metric.  Similarly
\cite{AlvarezGaumeFreedman} supersymmetry of a six-dimensional sigma
model with target space $X$ demands that $X$ be hyperK\"ahler.  Upon
dimensional reduction to one dimension these supersymmetric sigma
models become quantum mechanical systems, introduced to great effect
in \cite{WittenIndex} and much studied since.  These quantum
mechanical systems have as Hilbert space the square-integrable
differential forms on $X$, and as hamiltonian the Hodge laplacian.  By
reason of supersymmetry, the energy is non-negative.  Therefore the
ground states are in one-to-one correspondence with the
square-integrable harmonic forms and thus, when $X$ is compact, with
the cohomology.  Therefore any symmetry of the hamiltonian will
preserve the ground states and hence act on the cohomology.  On the
other hand, a $(d{+}1)$-dimensional supersymmetric sigma model with a
Lorentzian metric has an action which is invariant under the Lorentz
group $\SO(d,1)$.  Upon dimensional reduction to one (spatial)
dimension, this reduces to an $\SO(d{-}1,1)$ `internal' symmetry.
Applying this to supersymmetric sigma models in $3{+}1$
(resp. $5{+}1$) dimensions with compact target space $X$ yields an
$\SO(2,1)$ (resp. $\SO(4,1)$) action on the cohomology of $X$.  This
is Witten's argument.

Whereas this beautiful argument explains the existence of these group
actions, it does not tell us that they agree with the ones known from
geometry.  It is the purpose of this paper to prove that they do.  In
fact, supersymmetry lies at the heart of much of the Hodge--Lefschetz
theory and in this paper we will also briefly mention these results.
For example, the Hodge identities are what remains (after dimensional
reduction) of the fact that the supersymmetry generators are spinors of
the relevant (i.e., four- or six-dimensional) spacetime.

This paper is organised as follows.  In the next section we will
consider the K\"ahler case, reviewing the Hodge--Lefschetz
construction and then checking explicitly that it agrees with the one
coming from supersymmetry.  Then in section 3 we will tackle the
hyperK\"ahler case, reviewing Verbitsky's construction and checking
that it too agrees with that coming from supersymmetry.  Finally in
Section 4 we offer some closing comments.

\section{The K\"ahler case}

In this section we discuss the K\"ahler case.  We first briefly review
some facts from the Hodge--Lefschetz theory of a K\"ahler manifold which
we will need. We will then show how these may be understood
by considering the dimensional reduction of
the four-dimensional supersymmetric sigma model.
We follow the notational conventions of
\cite{WessBagger}.

\subsection{Some harmonic theory on K\"ahler manifolds}

Let $X$ be a compact K\"ahler manifold of complex dimension $n$.
The complex differential forms on $X$ decompose as
\begin{equation*}
A_c(X) = \bigoplus_{p,q=1,\ldots,n} A_c^{p,q}~,
\end{equation*}
where $A_c^{p,q}$ is the subspace of $(p,q)$-forms.
The Hodge $\star$-operator maps
\begin{equation*}
\star: A_c^{p,q} \to A_c^{n-q,n-p}~,
\end{equation*}
and obeys $\star^2 = (-)^{p+q}$ acting on $A_c^{p,q}$.  Using this
operator we may define an hermitian inner product on the space of
forms, the Hodge inner product:
\begin{equation}\label{eq:hodgeip}
(\alpha,\beta) = \int_X \alpha\wedge\star\Bar\beta~.
\end{equation}
Notice that because complex conjugation maps $A_c^{p,q} \to
A_c^{q,p}$, the Hodge inner product pairs forms of the same bidegree.

The de~Rham operator on $X$ decomposes into two operators of different
bidegree: $d = \d + \Bar\d$, where $\d : A_c^{p,q} \to A_c^{p+1,q}$
and $\Bar\d : A_c^{p,q} \to A_c^{p,q+1}$.  $\Bar\d$ is called the
Dolbeault operator.  From $d^2=0$, one reads that $\d^2 = \Bar\d^2 =
0$ and $\d\Bar\d = - \Bar\d\d$.  Let $\d^*$ and $\Bar\d^*$ denote
their adjoint operators relative to the Hodge inner product.  They
also satisfy similar identities.  Using these operators we can define
two laplacians $\square$ and $\Bar\square$ as follows:
\begin{equation*}
\square = \d\d^* + \d^* \d \qquad \text{and}\qquad 
\Bar\square = \Bar\d\Bar\d^* + \Bar\d^* \Bar\d~.
\end{equation*}
Part of the magic of K\"ahler manifolds is that these two laplacians
agree.  In fact, we have
\begin{equation}\label{eq:laplacians}
\square = \Bar\square = \tfrac{1}{2} \bigtriangleup~,
\end{equation}
where $\bigtriangleup$ is the Hodge laplacian.  Other identities
obeyed by the differential operators $\d$, $\Bar\d$, $\d^*$, and
$\Bar\d^*$ are:
\begin{equation}\label{eq:more-D-ids}
\d\Bar\d^* = - \Bar\d^*\d \qquad\text{and}\qquad 
\d^*\Bar\d = - \Bar\d\d^*~.
\end{equation}

Now let $\omega \in A_c^{1,1}$ denote the K\"ahler form.  It is real:
$\Bar\omega = \omega$.  Let $L: A_c^{p,q} \to A_c^{p+1,q+1}$ be
defined by $L(\alpha) = \alpha \wedge\omega$.  Its adjoint relative to
the Hodge inner product is denoted $\Lambda: A_c^{p,q} \to
A_c^{p-1,q-1}$, and obeys the relations
\begin{equation*}
(\Lambda(\alpha),\beta) = (\alpha, L(\beta)) = (\alpha, \beta \wedge
\omega)~.
\end{equation*}
It can be understood as contraction with the K\"ahler form.  Let us
now define a third operator $H \equiv [L,\Lambda] : A_c^{p,q} \to
A_c^{p,q}$.  Actually $H$ acts diagonally with real eigenvalue $p+q-n$
on $A_c^{p,q}$.  Furthermore, the following relations hold:
\begin{equation*}
{}[H,L] = 2L \qquad\text{and}\qquad [H,\Lambda] = -2\Lambda~.
\end{equation*}
In other words, $L$, $\Lambda$ and $H$ define a representation of the
Lie algebra $sl(2)$.

These operators obey a series of identities, known as the Hodge
identities, which relate the Dolbeault operator and its cousins.
These identities read:
\begin{alignat}{4}\label{eq:Hodgeids}
[L,\d] &= 0 &\quad [L,\Bar\d] &= 0 &\quad [\Lambda,\d^*] &= 0 &\quad
[\Lambda,\Bar\d^*] &= 0\notag\\
[L,\d^*] &= i\Bar\d &\quad [L,\Bar\d^*] &= -i\d &\quad [\Lambda,\d] &=
i\Bar\d^* &\quad [\Lambda,\Bar\d] &= -i\d^*~.
\end{alignat}
It follows from these identities, that $L$, $\Lambda$ and hence $H$
commute with the Hodge laplacian.  This means that they act on the
space of harmonic forms and, by the Hodge decomposition theorem, on
the cohomology.  As we will see below, all the identities in this
section will follow naturally from supersymmetry.

\subsection{The supersymmetric sigma model in four dimensions}

We will now recover these results using supersymmetry.  Let $X$ be a
K\"ahler manifold of complex dimension $n$.  Let $z^a$ and $\Bar
z^{\Bar a}$ denote local complex coordinates, relative to which the
metric has nonzero components $g_{a\Bar b}$.  The nonzero components
of the Christoffel symbols are denoted $\Gamma^a_{bc}$ and
$\Gamma^{\Bar a}_{\Bar b\Bar c}$, and those of the fully covariant
Riemann curvature tensor by $R_{a\Bar b c \Bar d}$.

The $N{=}1$ supersymmetric sigma model in four dimensions with target
space $X$ is described by the following lagrangian density:
\begin{equation}\label{eq:lagr}
\eL = -g_{a \Bar{b}} \d_{\mu} \phi^{a} \d^{\mu}
\Bar{\phi}^{\Bar{b}} - \tfrac{i}{2} g_{a \Bar{b}} \Bar{\chi}^{\Bar{b}}
\Bar{\sigma}^{\mu} \overleftrightarrow{D}_{\mu} \chi^{a} +
\tfrac{1}{4} R_{a \Bar{b} c \Bar{d}} (\chi^a\chi^c)
(\Bar\chi^{\Bar b}\Bar\chi^{\Bar d})~,
\end{equation}
where
\begin{itemize}
\item $\phi^a$ is a complex scalar field with $\Bar\phi^{\Bar a} =
(\phi^a)^*$;
\item $\chi^a_\alpha$ is a complex Weyl spinor and $\Bar\chi^{\Bar
a}_{\Dot\alpha} = (\chi^a_\alpha)^*$;
\item $\Bar\sigma^\mu = (-\1,-\sigma^i)$ and $\sigma^\mu =
(-\1,\sigma^i)$, where $\sigma^i$ are the Pauli matrices; and
\item the covariant derivative is defined by
\begin{equation*}
D_{\mu} \chi^{a} = \d_{\mu} \chi^{a} + \Gamma^{a}_{bc}
\d_{\mu} \phi^{b} \chi^{c}~.
\end{equation*}
\end{itemize}

The lagrangian \eqref{eq:lagr} is invariant under the superpoincar\'e
group, which contains an $\SO(3,1)$ subgroup generated infinitesimally
by:
\begin{align*}
\delta_{\Lambda} \phi^a &= - \Lambda_{\mu\nu} x^\mu
\d^{\nu}\phi^{a}\\
\delta_{\Lambda} \chi^a &= -\Lambda_{\mu\nu} x^\mu \d^\nu
\chi^a + \tfrac{1}{2} \Lambda_{\mu\nu} \sigma^{\mu\nu}\chi^{a}\\
\delta_{\Lambda} \Bar\chi^{\Bar a} &= -\Lambda_{\mu\nu} x^\mu
\d^\nu \Bar\chi^{\Bar a} + \tfrac{1}{2} \Lambda_{\mu\nu}
\Bar\chi^{\Bar a} \Bar\sigma^{\mu\nu}~,
\end{align*}
where $\Lambda_{\mu\nu}$ is constant and antisymmetric.  The Noether
current associated with $\so(3,1)$ transformations can be worked out
by letting the parameter $\Lambda_{\mu\nu}$ depend on position and
varying the lagrangian.
Then up to a total derivative, we find after some $\sigma$-algebra:
\begin{equation}\label{eq:current}
\delta_{\Lambda} \eL = \tfrac{i}{2} g_{a\Bar b} \d^\mu
\Lambda_{\mu\nu} \Bar\chi^{\Bar b} \Bar\sigma^\nu \chi^a~,
\end{equation}
where we have omitted terms which will not survive the dimensional
reduction.

The lagrangian \eqref{eq:lagr} is also invariant under the
supersymmetry transformations:
\begin{align*}
\delta_\varepsilon \phi^a &= \varepsilon\chi^a\\
\delta_\varepsilon \chi^a &= i\sigma^\mu \Bar\varepsilon \d_\mu\phi^a
- \Gamma^a_{bc}\delta_\varepsilon\phi^b \chi^c~,
\end{align*}
where $\varepsilon$ is a complex two-component spinor.  The
supersymmetry currents $S^\mu_\alpha$ and $\Bar S^{\mu\,\Dot\alpha}$
can be obtained just as was done for the Lorentz currents, and one
finds:
\begin{equation}\label{eq:susycurrent}
S^\mu = g_{a\Bar b} \d_\nu\Bar\phi^{\Bar b}
\sigma^\nu\Bar\sigma^\mu\chi^a~,
\end{equation}
with 
$\Bar S^{\mu\,\Dot\alpha} = \epsilon^{\Dot\alpha\Dot\beta}
\left(S^\mu_\beta\right)^*$.

We now perform a trivial dimensional reduction to one dimension by
simply dropping the dependence on all coordinates but $x^0$.  The
dimensionally reduced lagrangian then becomes:
\begin{equation*}
\eL = g_{a\bar b} \Dot\phi^a \Dot{\Bar\phi^{\Bar b}} - \tfrac{i}{2}
g_{a \bar b} \Bar\chi^{\Bar b} \bar\sigma^0
\overleftrightarrow{\frac{D}{dt}} \chi^a + 
\tfrac{1}{4} R_{a \Bar{b} c \Bar{d}} (\chi^a\chi^c)
(\Bar\chi^{\Bar b}\Bar\chi^{\Bar d})~,
\end{equation*}
with $\frac{D}{dt} = D_0$.  The conserved charges of the reduced
lagrangian are now
\begin{equation*}
J^{ij} = - {i\over2}g_{a\bar b}\bar\chi^{\bar
b}\bar\sigma^0\sigma^{ij}\chi^a,\qquad\text{with $i,j=1,2,3$,}
\end{equation*}
for the `internal' Lorentz generators, and
\begin{equation*}
S = g_{a\Bar b} \chi^a \Dot{\Bar\phi}^{\Bar b},
\end{equation*}
for the supercharges.

Defining $J_i = \tfrac{1}{2}\epsilon_{ijk}J^{jk}$ and after some
algebra, we find that
\begin{equation}\label{eq:noether}
J^i = \tfrac{i}{2} g_{a \Bar b} \Bar\chi^{\Bar b} \Bar\sigma^i
\chi^a.
\end{equation}

The quantisation of this system is well-known.  The quantisation of
the bosons $\phi^a$ and $\Bar\phi^{\Bar a}$ is straightforward: the
bosonic Hilbert space $\eB$ is the space of square integrable
complex-valued functions $f(\phi,\Bar\phi)$.  The canonical
anticommutation relations of the fermions are given by the following
Clifford algebra:
\begin{equation*}
\{\chi^a_\alpha, \Bar\chi^{\Bar b}_{\Dot\alpha}\} = -g^{a\Bar
b}(\phi,\Bar\phi)\sigma^0_{\alpha\Dot\alpha}~.
\end{equation*}
(Note that in our conventions $-\sigma^0_{\alpha\Dot\alpha}$ coincides
with the $2\times2$ identity matrix.)
We can choose a Clifford vacuum $|0\rangle$ by the condition that
$\Bar\chi^{\Bar b}_{\Dot2} |0\rangle = \chi^a_1 |0\rangle = 0$.  Then
a typical state in the fermionic Hilbert space $\eF$ is a linear
combination of monomials of the form:
\begin{equation*}
|\psi\rangle = \chi^{a_1}_2\chi^{a_2}_2\cdots\chi^{a_p}_2
\Bar\chi^{\Bar b_1}_{\Dot 1}\Bar\chi^{\Bar b_2}_{\Dot
1}\cdots\Bar\chi^{\Bar b_q}_{\Dot 1} |0\rangle~.
\end{equation*}
More precisely, states like these for fixed $p$ and $q$ generate a
subspace $\eF^{p,q}$ of the total fermionic Hilbert space.  The inner
product in $\eF$ is defined as follows: if $|\psi\rangle$ is as above,
its norm is given by $\langle\psi|\psi\rangle$ where
\begin{equation*}
\langle\psi| = \langle0| \chi_1^{b_q}\chi_1^{b_{q-1}}\cdots
\chi_1^{b_1} \Bar\chi_{\Dot2}^{\Bar a_p}\Bar\chi_{\Dot2}^{\Bar
a_{p-1}}\cdots\Bar\chi_{\Dot2}^{\Bar a_1}~.
\end{equation*}
Tensoring bosons and fermions together we see that the total Hilbert
space $\eH = \eB \otimes \eF$ is isomorphic to the space of
square-integrable complex differential forms on $X$ relative to the
Hodge metric.  Since $X$ is compact, $\eH$ has a dense subspace
isomorphic to the smooth complex forms $A_c(X)$.  Under the
isomorphism, $\eH$ inherits a bigrading $\eH = \bigoplus \eH^{p,q}$
which agrees with the one coming from the fermionic Hilbert space.

Under this isomorphism, operators in the quantum theory can be
interpreted geometrically as operators acting on $A_c(X)$.  It is not
hard to show that the following dictionary holds for the supersymmetry
generators:
\begin{equation}\label{eq:dictionary}
S_1 \mapsto \Bar\d^* \qquad S_2 \mapsto \d \qquad \Bar S_{\Dot1}
\mapsto \Bar\d \qquad \Bar S_{\Dot2} \mapsto \d^*~.
\end{equation}
The supersymmetry algebra obeyed by the supercharges, which can easily
be worked out by iterating the supersymmetry transformations, then
implies the identities \eqref{eq:laplacians} and
\eqref{eq:more-D-ids}.  In particular, this shows that up to an
inconsequential factor, the hamiltonian can be interpreted as the
Hodge laplacian.

Now consider the following linear combination of the Noether charges
\eqref{eq:noether}:
\begin{equation*}
L = J^{1} + i J^{2} = -i g_{a \Bar b}\Bar\chi^{\Bar b}_{\Dot1}\chi^a_2
= \omega_{a\Bar b}\Bar\chi^{\Bar b}_{\Dot1}\chi^a_2~.
\end{equation*}
Under the isomorphism $\eH \cong A_c(X)$, we see that $L$ agrees with
the operator $L$ defined in the previous subsection.  Its hermitian
adjoint is clearly given by
\begin{equation*}
\Lambda \equiv L^\dagger = - \omega_{b\Bar a}\Bar\chi^{\Bar a}_{\Dot2}\chi^b_1
= -(J^1 - i J^2)~.
\end{equation*}
Their commutator is given by
\begin{equation*}
H \equiv [L, L^\dagger] = 2 i J^3 = g_{a\Bar b} (\Bar\chi^{\Bar
b}_{\Dot 1} \chi_1^a - \Bar\chi^{\Bar b}_{\Dot 2} \chi_2^a)~,
\end{equation*}
which when written in terms of normal-ordered quantities becomes:
\begin{equation*}
H = g_{a\Bar b} (\Bar\chi^{\Bar b}_{\Dot 1} \chi_1^a +
\chi_2^a \Bar\chi^{\Bar b}_{\Dot 2}) - n\1~,
\end{equation*}
which acts as $(p+q-n)\1$ on $\eH^{p,q}$, as expected.  Thus under the
isomorphism $\eH \cong A_c(X)$, the operators $(L, \Lambda, H)$ go
over to their namesakes.  Notice that although the $J^i$ satisfy the
Lie algebra of $\so(3)$, the expressions for $L$, $\Lambda$ and $H$
involve complex linear combinations of the $J^i$.  The hermiticity
conditions are such that we are in effect choosing a different real
section of $\so(3,\CC)$, one isomorphic to $\so(2,1) \cong \sell(2)$.
Had we kept $x^1$, say, in the dimensional reduction, we would have
obtained $\so(2,1)$ directly without having to take complex linear
combinations; but such a choice would have made the notation a lot
more complicated.

\begin{table}[h!]
\begin{tabular}{|>{$}l<{$}|>{$}l<{$}|}
\hline
\multicolumn{1}{|c|}{Sigma model} & \multicolumn{1}{c|}{K\"ahler
geometry}\\
\hline\hline
\text{Hilbert space $\eH = \bigoplus_{p,q} \eH^{p,q}$} & A_c(X) =
\bigoplus_{p,q} A_c^{p,q} \\
\text{inner product} & \text{Hodge inner product \eqref{eq:hodgeip}}\\
\text{supercharge $\Bar S_{\Dot\alpha}$} & (\Bar\d,\d^*) \\
\text{supercharge $S_{\alpha}$} & (\Bar\d^*,\d) \\
\text{hamiltonian} & \bigtriangleup \\
\text{ground states} & H^*(X)\\
\text{`internal' symmetry $J^i$} & L,\,\Lambda,\,H \\
\text{supersymmetry algebra} &
\begin{cases}
\square = \Bar\square = \tfrac{1}{2} \bigtriangleup & \\
\text{$\d$-identities \eqref{eq:more-D-ids}} &
\end{cases}\\
\text{$(S_\alpha, \Bar S^{\Dot\alpha})$ is a spinor in $3{+}1$:
\eqref{eq:susyHodgeids}} & \text{Hodge identities
\eqref{eq:Hodgeids}}\\
\hline
\end{tabular}
\vspace{8pt}
\caption{Sigma models and K\"ahler geometry.\label{tab:susyklr}}
\end{table}

Finally we remark that the Hodge identities \eqref{eq:Hodgeids} are a
consequence of the spinorial nature of the four-dimensional
supercharge.  If we let $\tS = (S_\alpha~\Bar S^{\Dot\alpha})^t$
denote the four-dimensional supercharge, then $[J^{\mu\nu},\tS] =
\Sigma^{\mu\nu} \tS$.  Upon dimensional reduction, we find that
$[J^{ij}, S] = \sigma^{ij}\,S$, or simply that
\begin{equation}\label{eq:susyHodgeids}
[J^i,S] = \tfrac{i}{2} \sigma^i\Bar\sigma^0 S~.
\end{equation}
Using the expressions of $L$ and $\Lambda$ in terms of the $J^i$ and
the dictionary \eqref{eq:dictionary} it is easy to show that
\eqref{eq:susyHodgeids} goes over to the Hodge identities
\eqref{eq:Hodgeids}.  More invariantly, notice that $\tS$ is a
Majorana spinor in $3{+}1$ dimensions, hence under $\Spin^0(3,1) \cong
\SL(2,\CC)$, it transforms according to the real representation
$(0,\half) \oplus (\half,0)$.  Under the `internal' $\Spin^0(2,1)
\cong \SL(2,\RR)$ subgroup of $\Spin^0(3,1)$ this representation
breaks up into two doublets.  The Hodge identities simply reiterate
this fact. Table~\ref{tab:susyklr} above summarises the correspondence
described in this section.

\section{The hyperK\"ahler case}

Now comes the turn of the hyperK\"ahler case.  As before we first
briefly review Verbitsky's construction of the action of $\so(4,1)$ on
the cohomology of a compact hyperK\"ahler manifold and then show how
this can be reproduced via dimensional reduction of the
six-dimensional supersymmetric sigma model.

\subsection{Verbitsky's construction}

Let $X$ be a compact hyperK\"ahler manifold, and let $I$, $J$ and $K$
be the three complex structures and $\omega_i$, $i=1,2,3$ be the
corresponding K\"ahler forms.  We fix a choice of one of the
complex structures, $I$, say.  Ignoring for the moment the other two
complex structures, $X$ is a compact K\"ahler manifold and all of the
Hodge--Lefschetz theory goes through.  In particular we have operators
$L_1$, $\Lambda_1$ and $H$ defined as before but with $\omega_1$
playing the role of $\omega$.

We now bring to play the other two complex structures.  We define
operators $L_2$ and $L_3$ in the obvious way.  Their adjoints relative
to the Hodge inner product are $\Lambda_2$ and $\Lambda_3$.
Introducing operators $K_i = \tfrac{1}{2} \epsilon_{ijk}
[L_j,\Lambda_k]$, Verbitsky \cite{Verbitsky} showed that the
following algebra is satisfied:
\begin{alignat}{2}
[L_i, \Lambda_j] &= \epsilon_{ijk} K_k + \delta_{ij} H & \qquad
[K_i, K_j] &= \epsilon_{ijk} K_k \notag\\
[H, L_i] & = 2 L_i & \qquad [H, \Lambda_i] & = -2 \Lambda_i
\label{eq:so(4,1)}\\
[K_i,L_j] &= \epsilon_{ijk} L_k & \qquad
[K_i,\Lambda_j] &= \epsilon_{ijk} \Lambda_k\notag
\end{alignat}
with all other brackets zero.  Furthermore, Verbitsky also showed that
as in the K\"ahler case, these operators commute with the Hodge
laplacian, thus inducing an action on the cohomology.

The Lie algebra above is that of $\so(4,1)$. For the purposes of the
calculations in the next section, we will utilise the following
description of these generators.  Let $J_{mn}$, for $m,n =
1,\ldots,5$, denote the generators of $\so(5)$, satisfying the algebra
\begin{equation}\label{eq:so(5)}
[J_{mn} , J_{pq}] =  \delta_{mq} J_{np} - \delta_{mp} J_{nq} +
\delta_{np} J_{mq} - \delta_{nq} J_{mp}~,
\end{equation}
Now, defining (for $i,j = 1,2,3$)
\begin{alignat*}{2}
L_i &= -\left(J_{i5} + iJ_{i4}\right) &\qquad  \Lambda_i &= J_{i5} -
iJ_{i4},\\
H &= 2iJ_{45} &\qquad K_{ij} &= 2J_{ij}, 
\end{alignat*}
one finds that the algebra of these generators reproduces
\eqref{eq:so(4,1)}.  Again, taking complex linear combinations we have
moved to a different real section of $\so(5,\CC)$, this time
$\so(4,1)$ as evinced by the presence of factors of $i$ in connection
with the fourth coordinate.

\subsection{The six-dimensional supersymmetric sigma model}

We will now recover the results above using the supersymmetric sigma
model in $5{+}1$ dimensions.  The target space of this sigma model is
a hyperK\"ahler manifold $X$ of (real) dimension $4n$.  Normally the
fermions in the sigma model would be a section of the positive spinor
bundle $S_+$ over spacetime twisted by the pull-back of the tangent
bundle $T$ of the target manifold.  However in this case, this
prescription does not give rise to a match between the bosonic and
fermionic degrees of freedom: we must impose a restriction on the
fermions which we now detail.  The complexified tangent bundle $T_\CC$
of a hyperK\"ahler manifold decomposes under the maximal subgroup
$\Sp(1)\cdot \Sp(n) \subset \SO(4n)$ as $T_\CC \cong \Sigma \otimes
V$, where $\Sigma$ is a complex two-dimensional $\Sp(1)$ bundle and
$V$ is a complex $2n$-dimensional $\Sp(n)$ bundle.  The holonomy being
$\Sp(n)$ means that the above decomposition is preserved under
parallel transport and that in addition $\Sigma$ is a trivial bundle.
The canonical real structure of $T_\CC$ is the product of the natural
quaternionic structures in $\Sigma$ and $V$.  Because $S_+$ also
possesses a quaternionic structure, the tensor products $S_+\otimes V$
and $S_+\otimes\Sigma$ possess real structures.  Therefore we will be
able to impose reality conditions on the fermions and on the
supersymmetry parameters respectively.\footnote{Such spinors are known
as symplectic Majorana--Weyl spinors, and they exist in spacetimes of
signature $(s,t)$ with $s-t=4\mod 8$.}  The bundle $S_+\otimes V$ is
complex $8n$-dimensional.  The reality condition leaves $8n$ real
components which gives $4n$ physical degrees of freedom, matching the
number of bosonic physical degrees of freedom.  Similarly
$S_+\otimes\Sigma$ is complex 8-dimensional and the reality condition
leaves the expected 8 real components of the supercharge.

We now introduce some notation to describe the fields in the sigma
model.  First we have $4n$ bosons $\phi^i$ which are coordinates of the
target manifold.  The isomorphism $T_\CC \cong \Sigma\otimes V$ is
given explicitly by objects $\gamma^i_{A a}$.  Here
$A,B,\ldots$ are $\Sp(1)$ indices associated with $\Sigma$
and running from $1$ to $2$, and $a,b,\ldots$ are $\Sp(n)$ indices
associated with $V$ and running from $1$ to $2n$.  The bundle $\Sigma$
being trivial allows a constant $\Sp(1)$-invariant symplectic form
$\epsilon_{AB}$; whereas $V$ admits an $\Sp(n)$-invariant symplectic
form $\omega_{ab}$.  In terms of these symplectic forms, the metric
$g_{ij}$ on $T$ can be written as:
\begin{equation*}
g_{ij} \gamma^i_{A a} \gamma^j_{B b} = \epsilon_{AB}
\omega_{ab}~.
\end{equation*}
Because the holonomy lies in $\Sp(n)$, not just the metric $g$, but
also the symplectic forms $\epsilon$ and $\omega$ are parallel; whence
so are the maps $\gamma^i_{A a}$.  We choose to trivialise $\Sigma$
globally and put on it the zero connection.  This way any constant
section is parallel.

A final piece of notation is to choose an explicit realisation for the
Clifford algebra in $5{+}1$ dimensions (this will determine the
explicit form of the reality condition satisfied by the fermions).  
The metric is $\eta_{mn} = {\rm diag}(-1,1,1,1,1,1)$.
We choose the ($5+1$)--dimensional Gamma matrices to be
\begin{equation*}
\Gamma_\mu = \begin{pmatrix}
0 & \gamma_\mu \\
\gamma_\mu & 0
\end{pmatrix}\quad,\quad
\Gamma_4 = \begin{pmatrix}
0 & -\1 \\ \1 & 0
\end{pmatrix}\quad\text{and}\quad
\Gamma_5 = \begin{pmatrix}
0 & \gamma_5 \\ \gamma_5 & 0 
\end{pmatrix}~,
\end{equation*}
where $\gamma_\mu$, for $\mu=0,1,2,3$, are the ($3{+}1$)-dimensional
gamma matrices:
\begin{equation*}
\gamma^{\mu} = \begin{pmatrix} 0 & \sigma^\mu \\ \Bar\sigma^\mu & 0
\end{pmatrix}\qquad\text{and}\qquad \gamma_5 =
\gamma_0\gamma_1\gamma_2\gamma_3~.
\end{equation*}
Finally, we choose $\Gamma_7 =
\Gamma_0\Gamma_1\Gamma_2\Gamma_3\Gamma_4\Gamma_5$.

We can now write down the following lagrangian (see, e.g.,
\cite{RozanskyWitten}):
\begin{equation*}
\eL = \tfrac{1}{2} g_{ij} \d_m \phi^i \d^m \phi^j + \tfrac{1}{2}
\omega_{ab} \Bar\Psi^a \Gamma^m D_m \Psi^b - \tfrac{1}{48}
\Omega_{abcd} (\Bar\Psi^a \Gamma_m \Psi^b) (\Bar\Psi^c
\Gamma^m\Psi^d)~.
\end{equation*}

In this expression,
\begin{itemize}
\item the $\Psi^a$ are eight-component positive-chirality Weyl
spinors: $\Gamma_{7} \Psi^a = + \Psi^a$.  In the above basis for the
$\Gamma$-matrices, it means that $\Psi^a =
\left(\begin{smallmatrix}\psi^a \\ 0 \end{smallmatrix}\right)$, with
$\psi^a$ a four-component complex Dirac spinor, $\psi^a =
\left(\begin{smallmatrix} \chi^a_{\alpha} \\
\Bar{\phi}^{a\,\Dot{\alpha}}\end{smallmatrix}\right)$.
\item the $\Psi^a$ {\it also} satisfy the following symplectic
Majorana condition:
\begin{equation*}
\Psi_a^* \equiv \left(\Psi^a\right)^* = \omega_{ab} \sB \Psi^b~,
\end{equation*}
where the matrix $\sB$ must satisfy \cite{KugoTownsend} $\Gamma_m^* =
\sB\Gamma_m \sB^{-1}$, $\sB^t=-\sB$ and $\sB^\dagger \sB= \1$.  In our
choice of basis we can take $\sB=\Gamma_2\Gamma_5$.  This condition
relates further the two-component spinors comprising $\psi^a$ in such
a way that $\psi^a = \left(\begin{smallmatrix} \chi^a_{\alpha} \\ -
\omega^{ab} \epsilon^{\Dot\alpha\Dot\beta}
\Bar\chi_{b\Dot\beta}\end{smallmatrix}\right)$, where
$\Bar\chi_{b\Dot\beta} \equiv \left( \chi^b_{\beta} \right)^*$.
\item the conjugate $\Bar\Psi^a$ is given by $\Bar\Psi^a = (\Psi^a)^t
\sC$, where $\sC$ is the charge conjugation matrix, which in our basis
will be chosen to be $\sC = -\Gamma_2\Gamma_5\Gamma_0$;
\item the covariant derivative is given by
\begin{equation*}
D_m \Psi^a = \d_m \Psi^a + \Hat\Gamma_{i~b}^a \d_m\phi^i \Psi^b~,
\end{equation*}
with $\Hat\Gamma$ the reduction to $\Sp(n)$ of the riemannian
connection; and
\item $\Omega_{abcd}$ is the hyperK\"ahler curvature, a totally
symmetric tensor defined by
\begin{equation*}
\gamma^i_{A a} \gamma^j_{B b} \gamma^k_{C c}
\gamma^\ell_{D d} R_{ijk\ell} = \epsilon_{AB}\epsilon_{CD}
\Omega_{abcd}~.
\end{equation*}
\end{itemize}

The above lagrangian is invariant under supersymmetry transformations:
\begin{align*}
\delta_\varepsilon \phi^i &= \gamma^i_{A a} \Bar\varepsilon^A\Psi^a\\
\delta_\varepsilon \Psi^a &= \gamma_i^{A a} \Gamma^m \d_m \phi^i
\varepsilon_A - \Hat\Gamma_{i~b}^a \delta_\varepsilon\phi^i \Psi^b
\end{align*}
where $\varepsilon^A$ is a constant {\em negative\/}-chirality Weyl
spinor with values in $\Sigma$ subject to the symplectic Majorana
condition:
\begin{equation*}
\varepsilon_A^* \equiv \left(\varepsilon^A\right)^* = \epsilon_{AB}
\sB \varepsilon^B~,
\end{equation*}
and $\Bar\varepsilon^A = \left(\varepsilon^A\right)^t \sC$.  The
Noether current generating the supersymmetry is given by
\begin{equation*}
S^{Am} = \omega_{ab} \gamma_i^{Ab} \d_n\phi^i\Gamma^n\Gamma^m\Psi^a~.
\end{equation*}

The above lagrangian is also invariant under the following
infinitesimal Lorentz transformations satisfying the Lie algebra of
$\so(5,1)$:
\begin{align*}
\delta_\Lambda \phi^i &= - \Lambda_{mn} x^m \d^n \phi^i\\
\delta_\Lambda \Psi^a &= -\Lambda_{mn} x^m \d^n \Psi^a + \tfrac{1}{2}
\Lambda_{mn} \Sigma^{mn}\Psi^a~.
\end{align*}
As in the four-dimensional sigma model, we compute the Noether
current by letting $\Lambda_{mn}$ depend on the position and varying
the lagrangian. The time (zeroth) component of this current, whose
integral over space yields the conserved quantities, is found to be
\begin{equation}\label{eq:hkic}
J_{mn} = \tfrac{1}{4} \omega_{ab} \Bar\Psi^a\Gamma^0
\Gamma_{mn}\Psi^b,~
\end{equation}
where $\Gamma_{mn} = \tfrac{1}{2}(\Gamma_m\Gamma_n -
\Gamma_n\Gamma_m)$, and where, in anticipation, we have omitted terms
which will not survive the dimensional reduction.

We now retain $x^0$ as the time and drop all dependence on the other
coordinates.  It is then a simple matter to impose the symplectic
Majorana Weyl condition upon the fermions and use the explicit
realisation for the $\Gamma$-matrices, to derive dimensionally reduced
expressions for the quantities of interest.  We will do this later for
the `internal' symmetry generators.

Finally, one finds after a little calculation that the term in the
Lagrangian density which is quadratic in the fermions may be written
$i\bar\chi_a \Bar\sigma^0\Dot\chi^a$.  After quantisation, one thus
finds the anticommutation relations amongst the fermions
\begin{equation*}
\{\chi^a_\alpha, \Bar\chi_{b\Dot\alpha} \} = - \delta^a_b
\sigma^0_{\alpha\Dot\alpha}~,
\end{equation*}
where again we remind the reader that in our conventions, $\sigma^0 =
-\1$.
We will choose the $\chi^a_\alpha$ as creation operators and the
$\bar\chi_{a\Dot\alpha}$ as annihilation operators, acting on the
appropriate Clifford vacuum.  The bosons are quantised in the usual
way, their Hilbert space being the space of square-integrable
functions $f(\phi)$.  Just as in the K\"ahler case, the total Hilbert
space is (the completion of) the space of smooth complex-valued
differential forms $A_c(X)$ on the hyperK\"ahler manifold $X$.  The
explicit map is the following:
\begin{equation}\label{eq:geoisohk}
f_{i_1i_2\cdots i_k}(x) dx^{i_1}\wedge dx^{i_2}\wedge \cdots\wedge
dx^{i_k} \leftrightarrow f_{i_1i_2\cdots i_k}(\phi) \chi^{i_1}
\chi^{i_2} \cdots \chi^{i_k}|0\rangle~,
\end{equation}
where
\begin{equation*}
\chi^i \equiv \gamma^i_{a1} \chi^{a1} + \gamma^i_{a2} \Bar\chi^{a2}~.
\end{equation*}
Notice that in this equation we have identified the index $A$
pertaining to $\Sigma$ with the index $\alpha$.  The inner product is
defined as follows: if $|\psi\rangle = \chi^{a_1}_{\alpha_1}
\chi^{a_2}_{\alpha_2} \cdots \chi^{a_k}_{\alpha_k} |0\rangle$, its
norm is given by $\langle\psi|\psi\rangle$ where
\begin{equation}\label{eq:hknorm}
\langle\psi| = \langle0|
\Bar\chi_{a_k\Dot\alpha_k} \Bar\chi_{a_{k-1}\Dot\alpha_{k-1}} 
\cdots\Bar\chi_{a_1\Dot\alpha_1}~.
\end{equation}
In other words, just as in the K\"ahler case the adjoint is complex
conjugation: $\left(\chi^a_\alpha\right)^\dagger =
\Bar\chi_{a\Dot\alpha} = \left(\chi^a_\alpha\right)^*$.  We see that
again the Hilbert space of the quantum mechanical sigma model is (the
completion of) the space of complex-valued smooth differential forms
relative to the Hodge inner product.  It is now possible to quantise
the supercharges and compute their algebra. As is well-known, one
finds that the hamiltonian agrees (up to a factor) with the Hodge
laplacian.  Therefore the ground states are once again in one-to-one
correspondence with the cohomology.

As in the K\"ahler case, it is possible to make a dictionary relating
the quantities appearing in the sigma model with the geometry of $X$.
However we will simply mention that again the supercharges go to
differential operators.  Choosing a complex structure on $X$, it is
possible to single out a particular linear combination of the
supercharges which can be identified with the Dolbeault operator.  In
fact, on $X$ there is a 2-sphere's worth of complex structures, on
which the Lie group $\SO(3)$ acts transitively.  On the sigma model
side, this is nothing but the R-symmetry of the six-dimensional
supersymmetry algebra.  There are also analogues of the Hodge
identities.  Just as in the K\"ahler case, these identities reflect
the transformation properties of the supercharges under the `internal'
symmetry subgroup $\Spin^0(4,1) \cong \Sp(1,1)$ of $\Spin^0(5,1) \cong
\SL(2,\HH)$.  The supercharges, being Weyl spinors, transform as the
canonical irreducible representation of $\SL(2,\HH)$ of quaternionic
dimension 2, which under the `internal' $\Sp(1,1)$ subgroup remains
irreducible.

We now come to the main point of this section: the
supersymmetric origins of the action of $\so(4,1)$ on $H^*(X)$.
As before, this action will come induced from the action of the
Lorentz generators under the reduction to one dimension. The
generators $J_{mn}$, for $m,n = 1,...,5$, now generate an `internal'
$\so(5)$ symmetry of the one-dimensional supersymmetric quantum model
obtained by this reduction.  By virtue of its six-dimensional origin,
this symmetry will furthermore commute with the action of the
Hamiltonian, hence will induce an action on the ground states, i.e.,
on the cohomology of $X$.  It remains to show that out of this
$\so(5)$ symmetry follows the $\so(4,1)$ symmetry introduced by
Verbitsky and  summarised in the previous section.  We do so now.

It is enough to consider the linear combinations $L_{i} = -(J_{i5} + i
J_{i4})$.  Expanding the expression \eqref{eq:hkic}, one finds that
\begin{align}
L_i &= -\tfrac12 \omega_{ab} (\chi^a)^t \sigma^2\sigma^i\chi^b\notag\\
\intertext{or more explicitly}
L_1 &= \phantom{-}\tfrac{i}{2} \omega_{ab} \left( \chi^a_1\chi^b_1 - 
 \chi^a_2\chi^b_2 \right)\notag\\
L_2 &= -\tfrac12 \omega_{ab} \left( \chi^a_1\chi^b_1 +
 \chi^a_2\chi^b_2 \right)\notag\\
L_3 &= -\tfrac{i}{2} \omega_{ab} \left( \chi_1^a \chi_2^b + \chi_2^a
\chi_1^b \right)~.\label{eq:Liexplicit}
\end{align}
We define $\Lambda_i$ as the adjoints of $L_i$ with respect to the
inner product defined by the norm \eqref{eq:hknorm}, and then the
other operators are defined using the Lie algebra of $\so(4,1)$. In
other words, $K_i$ and $H$ are defined by $K_i = \tfrac12
\epsilon_{ijk}[L_j,\Lambda_k]$, and $H=[L_1,\Lambda_1]$, say.  One can
then check that these operators satisfy the Lie algebra $\so(4,1)$ as
written in \eqref{eq:so(4,1)}.

To conclude the proof we must show that this $\so(4,1)$ agrees with
the one introduced by Verbitsky.  Clearly it is sufficient to check
that the $L_i$ can be interpreted as exterior product with the three
K\"ahler forms.  Under the isomorphism \eqref{eq:geoisohk}, $L_i$
indeed corresponds to exterior product with a 2-form $\omega_i \in
\bigwedge^2T^* \cong \bigwedge^2(\Sigma\otimes V)$ with components
$\omega^i_{aAbB} = \omega_{ab} M^i_{AB}$, where $(M^i)^{AB}$ are
matrices which can be read off from \eqref{eq:Liexplicit}:
\begin{equation*}
M^1 = \begin{pmatrix}i & 0 \\ 0 & -i\end{pmatrix}\quad
M^2 = \begin{pmatrix}-1 & 0 \\ 0 & -1\end{pmatrix}\quad
M^3 = \begin{pmatrix}0 & -i \\ -i & 0\end{pmatrix}~.
\end{equation*}
It is clear that the forms $\omega^i$ are parallel, since so are
$\omega$ and any constant section of any power of $\Sigma$ (recall
that $\Sigma$ is a trivial bundle with the zero connection).  It is
moreover clear that these forms are linearly independent.  Since in an
otherwise arbitrary compact hyperk\"ahler manifold $X$ the space of
parallel 2-forms is three-dimensional and spanned by the K\"ahler
forms, we conclude that these are the K\"ahler
forms.  However we can also
explicitly write down the complex structures in this basis.  The
complex structures are defined by $(I^i)^{aA}_{bB} =  \omega^i_{cCbB}
= \delta^a_b (J^i)^A_B$, where the matrices $J^i$ are related to the
$M^i$ by $(J^i)^A_B = (M^i)^{AC} \epsilon_{CB}$:
\begin{equation*}
J^1 = \begin{pmatrix}0 & i \\ i & 0 \end{pmatrix}\quad
J^2 = \begin{pmatrix}0 & -1\\ 1 & 0 \end{pmatrix}\quad
J^3 = \begin{pmatrix}i & 0 \\  0 & -i\end{pmatrix}~.
\end{equation*}
It is easy to see that they obey the algebra of the imaginary
quaternions:
\begin{equation*}
J^i\,J^j = - \delta^{ij} \1 + \epsilon^{ijk} J^k~,
\end{equation*}
as expected.

\section{Conclusions}

In this paper we have shown, following an idea of Witten, how the
Hodge--Lefschetz theory for (hyper)K\"ahler manifolds is a natural
consequence of supersymmetry.  That such a statement can be made
should not come as a surprise because K\"ahler and hyperK\"ahler
manifolds can be {\em defined\/} by the fact that it is 
only on these manifolds that the
($3{+}1$)- and ($5{+}1$)-dimensional sigma models admit
supersymmetry.  That this is the case is also linked in a special way
to the properties of Clifford algebras in those dimensions: Weyl
spinors in ($3{+}1$) dimensions are complex, whereas in ($5{+}1$)
dimensions they are quaternionic.  Furthermore, that the
dimensionally reduced sigma model should have anything to do with the
Hodge--Lefschetz theory (or Verbitsky's extension) is again not
surprising, since this theory deals with the cohomology of the
manifold, which are the quantum-mechanical ground states of the
one-dimensional sigma model.

It is tempting to speculate that other aspects of the geometry of
these manifolds also have supersymmetric origins.  For example, if $X$
is compact hyperK\"ahler, Verbitsky \cite{Verbitsky} proved that in
general, there is an action of $\so(4,b_2-2)$ on $H^*(X)$, where $b_2
= \dim H^2(X)$.  It remains an open problem to find a supersymmetric
origin to this symmetry.

\begin{ack}
We would like to acknowledge useful conversations with Jerome
Gauntlett and Chris Hull.  In addition, one of us (JMF) would like to
thank the Isaac Newton Institute for Mathematical Sciences of the
University of Cambridge for its hospitality and support during the
early and final stages of this work.
\end{ack}


\begin{thebibliography}{10}

\bibitem{AlvarezGaumeFreedman}
L~Alvarez-Gaum\'e and DZ~Freedman,{\em Geometrical structure and
ultraviolet finiteness in the supersymmetric sigma model},
\CMP{80}{1981}{443}.

\bibitem{GriffithsHarris}
P~Griffiths and J~Harris, {\em Principles of algebraic geometry},
Wiley 1978.

\bibitem{KugoTownsend}
T~Kugo and P~Townsend , {\em Supersymmetry and the division algebras}, 
\NPB{221}{1983}{357-380}.

\bibitem{RozanskyWitten}
L~Rozansky and E~Witten, {\em HyperK\"ahler geometry and
invariants of three manifolds}, {\tt hep-th/9612216}.

\bibitem{Verbitsky-so(5)}
M~Verbitsky, {\em On the action of a Lie algebra $\so(5)$ on the
cohomology of a hyperk\"ahler manifold}, \FAA{24(2)}{1990}{70--71}.

\bibitem{Verbitsky}
M~Verbitsky, {\em Cohomology of compact hyperK\"ahler manifolds},
{\tt alg-geom/9501001}.

\bibitem{WessBagger}
J~Wess and J~Bagger, {\em Supersymmetry and supergravity}, Princeton
1983.

\bibitem{WittenIndex}
E~Witten, {\em Constraints on supersymmetry breaking},
\NPB{202}{1982}{253-316}.

\bibitem{WittenTalk}
E~Witten, talk given at the Newton Institute for Mathematical
Sciences, Cambridge, November 1996.

\bibitem{Zumino}
B~Zumino, {\em Supersymmetry and K\"ahler manifolds},
\PLB{87}{1979}{203}.

\end{thebibliography}
\end{document}